\documentclass[11pt]{article}
\usepackage{epsfig}
\usepackage{natbib}

\def\etal{{et~al.}}
\def\farcs{\hbox{$.\!\!^{\prime\prime}$}}
\def\received#1{\vspace*{-1.5ex}{\topsep\z@\center{Received #1}\endcenter}}
\def\accepted#1{\vspace*{-1.5ex}{\topsep\z@\center{Accepted #1}\endcenter}}
\def\affilmark#1{$^{#1}$}
\sloppypar

\begin{document}

\newcommand{\ups}{\rule{0pt}{15pt}}

\title{\bf 
Measurement of optical turbulence in free atmosphere above Mt.Maidanak in
2005-2007
}
\date{Received 26.12.2008}
\author{ \bf \copyright\, 2008. \ \ V.\,Kornilov\affilmark{1*},
S.\,Ilyasov\affilmark{2}, O.\,Voziakova\affilmark{1}, Yu.\,Tillaev\affilmark{2}, \\ \bf
 B.\,Safonov\affilmark{1}, M.\,Ibragimov\affilmark{2}, N.\,Shatsky\affilmark{1},
S.\,Ehgamberdiev\affilmark{2} \\
$^1${\it Sternberg astronomical institute, Universitetskiy prosp. 13, Moscow,
Russia}\\
$^2${\it Ulughbek Astronomical institute, ul. Astronomicheskaya, 33,
Tashkent, Uzbekistan}
}
\maketitle


\hspace{1.5cm} Accepted for publication in Astronomy Letters, Volume 35, 2009

\sloppypar 
\vspace{2mm}
\noindent
\begin{abstract}
\large  Results of 2005-2007  campaign of  measurement of  the optical turbulence    vertical    distribution    above    Mt.~Maidanak    are presented.  Measurements are performed  with the  MASS (Multi-Aperture Scintillation  Sensor)  device  which  is widely used  in  similar studies  during  last  years   at  several  observatories  across  the world. The data  analysis shows that median seeing  in free atmosphere (at altitudes above 0.5km) is $0\farcs46$ and median isoplanatic
angle  is   $2\farcs47$.   Given  a   rather  long  atmospheric coherence time (about 7~ms when the seeing is good) such conditions are favorable for adaptive optics and interferometry in the visible and near-IR.
\end{abstract}
\noindent
\hspace{1.5cm}{\bf Key words:} astroclimate, seeing, adaptive optics


\vfill
\noindent\rule{8cm}{1pt}\\
{$^*$ e-mail: $<$victor@sai.msu.ru$>$}


\section{INTRODUCTION}

Studies  of  atmospheric  optical turbulence  have  progressed significantly in the last  decade. This was stimulated by 
new programs of site  testing for extra-large telescopes. It was also realized that the  potential of existing observatories  may be further enhanced by the use of adaptive optics (AO) systems.

It is well known that knowledge of only the integral turbulence on the line of sight is not enough for correct prediction of AO efficiency at a given  site. Information on the vertical  distribution of turbulence is needed  for the development of  a particular kind  of AO \citep[see for  example,][]{vern1991,Rodierbook, Wils2003, TPcan,  CTIO03}.  It is  also well  known  that atmospheric limits on  photometric and  astrometric precision  are directly linked  to the  vertical profile  of turbulence
\citep{drav97, ken2006, shao}.  It is worth  noting that the discussion of this issue  in the 1970--80th was  mainly related to the selection  of optimal height of telescope towers.

Astroclimate parameters of Maidanak  observatory were studied for many years. One such  campaign was conducted in 1998--99, as   reported by Ilyasov et  al.  \citeyearpar{I99} and Ehgamberdiev et al \citeyearpar{eso}. First estimates of the contribution of  free atmosphere to  seeing were  made by \citep{Free} based on stellar scintillation analysis.  It turned out  that  this contribution was quite significant, about 30\%.

The development of an effective technique to study vertical turbulence distribution by stellar scintillations \citep{Marr, MASS, Rest} and of respective instrumentation allowed us to  initiate in 2005 the project of  turbulence  monitoring  at  Mt.~Maidanak during 2--3  years, completed in 2007. Here we report its main results. 

\section{MEASUREMENT METHOD AND MULTI-APERTURE SCINTILLATION SENSOR (MASS)}

The  measurement of  the altitude distribution optical  turbulence is based on the fact that  amplitude of stellar scintillation produced by a turbulent  layer depends on its  height in a different  way when observed through different apertures. 

As  well  described  in  \cite{Ta67, R81},  stellar  scintillation  is produced  by  phase  fluctuations  of  a light  wave  passing  through optically turbulent layers  of atmosphere, and subsequent propagation. The strength  and spatial spectrum of  amplitude fluctuations produced by the atmosphere depend on the propagation length.

Scintillation   is  quantified  by   the  scintillation   index  $s^2$ representing  the relative variance of flux in some aperture. Tokovinin (\citeyear{To98}) generalizes this concept to the  case  of simultaneous  measurement  of  fluxes  in two  different apertures.   The  two kinds  of  scintillation  indices  -- so  called ``normal'' and differential -- can be directly measured.  On the other hand,  these indices  equal integrals  of the  turbulence distribution along the line of sight with some weighting functions. The weights are computed   from  the   known  aperture   sizes  and   spectral  energy
distribution of the  light. It should be noted  that the validity of the small-perturbations  method   and  of  the  Kolmogorov  turbulence spectrum   is  essential   here.  In   typical  conditions   of  astronomical observations, both these assumptions  are usually valid.

Further development of this method  is described in papers by Kornilov et al. (\citeyear{MASS}) and Tokovinin et al. (\citeyear{Rest}), where specific algorithms  of restoration of turbulence along  the line of sight  are presented  and various  instrumental factors  are accounted for. As  dedicated studies show \citep{TK07, Kor07},  the precision of
turbulence strength representation by a set of six altitude layers is definitely better than 10\%.  An independent cross-check of these results at some observatories \citep{RMex2007,Lasilla2007}  was  made  with a  SCIDAR  device
\citep[Scintillation detection and ranging,][]{sciorg,Fuchs}.

\begin{table}[h]
\caption{MASS entrance aperture diameters ( $d_I$ --- inner diameter, $d_O$ --- outer diameter).\label{tab:1}}
\bigskip
\centering
\begin{tabular}{p{2cm}p{2cm}p{2cm}p{2cm}p{2cm}}
\hline\hline
Channel & \multicolumn{2}{l}{Segmentator mirrors} &\multicolumn{2}{l}{Effective~aperture}\\
 MASS     & $d_I$, mm & $d_O$, mm & $d_I$, cm & $d_O$, cm \\[5pt]
\hline
A     & \ups---   & 1.03 & ---  & 1.60  \\
B     & 1.05      & 1.90 & 1.62 & 2.93  \\
C     & 1.95      & 3.60 & 3.00 & 5.54  \\
D     & 3.65      & 6.75 & 5.62 & 10.4  \\[5pt]
\hline\hline
\end{tabular}
\end{table}

The  MASS  device  is simply a   fast  4-channel photometer which measures fluctuations of the light flux from
some reasonably  bright star  in four concentric  apertures from  2 to 10~cm  diameters.  Detailed description  of  the  device  is given  by Kornilov et al  (\citeyear{MASS}). Any 5- to 6-inch telescope of about 3m focal  distance and  without central obscuration  may be used  as a feeding optics.

A  natural  drawback  of  the  MASS method  is  its  insensitivity  to turbulence  near the  ground. In  order  to include  it for  accessing complete  turbulence integral  along  the line  of  sight (giving  the seeing  measure),  one has  to  use  some  other method,  for  example differential image motion  monitor \citep[DIMM,][]{DIMM}.  It is worth recalling  here that  {\it image  quality  (seeing)} is  defined as  a full-width at half-maximum of a long-exposure stellar image created by
a large aberration-free telescope.  Seeing produced by the whole atmosphere without a contribution  from  the ground  layer  is  referred  to as  {\it  free atmosphere  seeing}  $\beta_{free}$.    A reliable  estimate  of  this quantity is normally obtained for turbulence above $h = 0.5 \mbox{km}$ \citep{MASS, Rest}.

The altitude distribution of turbulence above Mt.~Maidanak was studied with help of so-called {\it  original} MASS device of first generation which  was   designed  and  built  in  2001--2002   in  the  Sternberg Astronomical  Institute in  collaboration  with the  Cerro Tololo  and European Southern observatories.

\section{MAKING MEASUREMENTS AND SUPPLEMENTARY STUDIES}
\begin{figure}[!h]
\center
\psfig{figure=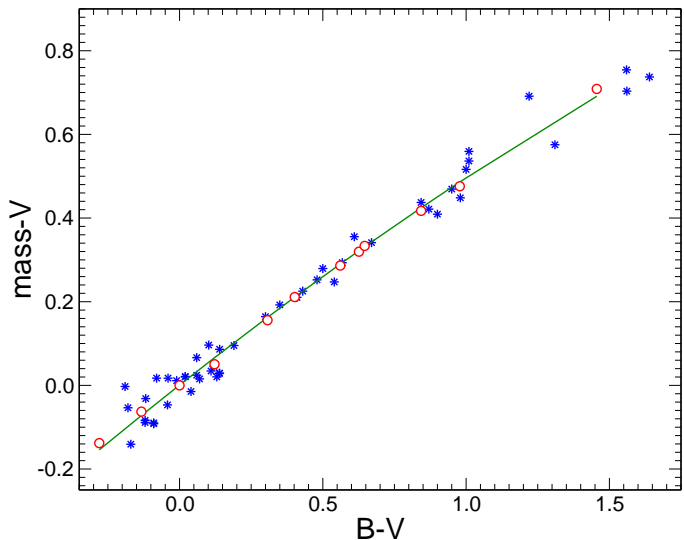,height=10cm,angle=-90}
\caption{ Colour equation of MASS at the refractor telescope AFR-2. Asterisks -- real stars, a solid curve -- parabolic approximation, circles -- computed colour dependencies for a set of used typical spectral energy distributions after the applied response curve correction.  \label{fig:sp}}
\end{figure}

The observations with the MASS were made from August 2005 to November 2007. The device was attached to the photographic refractor telescope AFR-2 of 23~cm aperture and f/no 10. Valid data are only obtained after a proper adjustment of the Fabry lens which sets the sizes of effective work apertures projected onto the telescope entrance aperture plane \citep{MASS, Kor07}. The respective magnification factor equal to 15.4 was measured after installation and was confirmed by repetitive measurements in 2006. The physical sizes of the MASS segmentator and the effective apertures of respective channels are given in Table~\ref{tab:1}. One can see that aperture size for the channel~A is smaller than a typical Fresnel radius $\sqrt{\lambda H}$ = 3 -- 7~cm depending on turbulence height while the D-channel aperture is larger which is a must for successful application of profiles restoration algorithms.

The weighting functions converting the turbulence distribution along the line of sight into scintillation indices depend on spectrum of light. Correct account of this factor is implemented using typical energy distributions in spectra of program stars and carefully the measured response curve of the device. While the device itself is well studied in a laboratory, the refractor lens introduces significant absorption in the blue and hence strongly violates the actual response. Correction of this effect was made using photometric measurements of some 50 stars taken 15 and 16 October 2006.

\begin{figure}[h]
\center
\psfig{figure=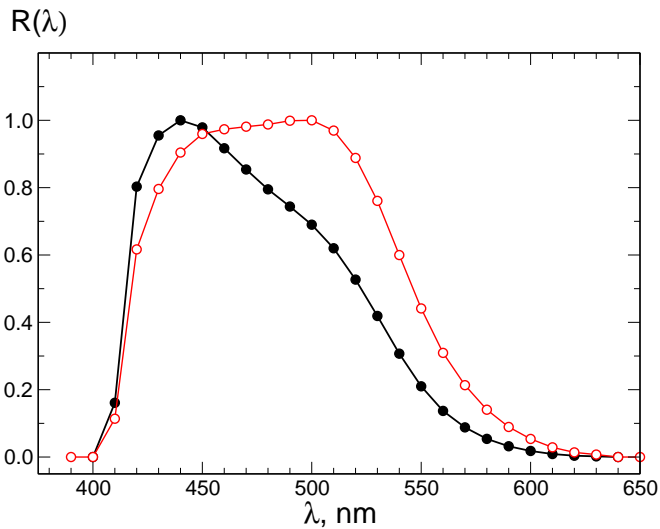,height=10cm,angle=-90}
\caption{ Original (black dots) and corrected (empty circles) spectral response curves of a MASS $+$ telescope system.  \label{fig:resp}}
\end{figure}

Using the obtained atmospheric extinction coefficient ($0^m\!.27$ in the MASS bandpass), the colour equation  $mass-V$ vs $B-V$ was constructed (Fig.~\ref{fig:sp}). The slope of this relation allowed us to correct the response of the system given an assumed typical glass selective absorption law\footnote{See Kornilov V., {\it The verification of the MASS spectral response}.\\
{\tt http://dragon.sai.msu.ru/mass/download/doc/mass\_spectral\_band\_eng.pdf}, 2006.}. The original MASS response curve and those after correction are shown in Fig.~\ref{fig:resp}.

Unbiased estimation of turbulence strengths, especially when it is weak, is only possible given the correctly measured detector non-poissonity factors $p$ in A and B channels. The flux in C and D apertures is normally strong enough to neglect this effect. Non-poissonity measurements involve the control light source which is built into the device \citep{Kor07, TK07}. Control measurements give also the non-linearity (dead-time) measure $\tau$ which, contrary to $p$, is most important for C and D channels where bright stars may develop more than $10^6$ counts per second. The results of these supplementary measurements made in two separate nights of 2006 are given in Table~\ref{tab:2}.

\begin{table}
\caption{The device parameters  $p$ (non-poissonity) and  $\tau$ (non-linearity) by channels measured in 2006. \label{tab:2}}
\bigskip
\centering
\begin{tabular}{p{2cm}p{2cm}p{2cm}p{2cm}p{2cm}}
\hline\hline
Channel& $p$ & $\sigma_p$ & $\tau$, ns & $\sigma_\tau$, ns\ups \\[5pt]
\hline
A \ups& 1.010     & 0.001 & 22.8 & 3.5  \\
B     & 1.012     & 0.001 & 21.4 & 1.4  \\
C     & 1.008     & 0.001 & 21.0 & 0.3  \\
D     & 1.007     & 0.001 & 19.4 & 0.2  \\[5pt]
\hline\hline
\end{tabular}
\end{table}
The statistics of observation time distribution by seasons is given in Table~\ref{tab:3}. A total of 280 nights was occupied by measurements of which a small fraction (25 nights) were instances of short estimates of less than half an hour duration. Total measurement time is 1022 hours. The gap in June -- September 2006 is due to device temporal failure.

\begin{table}
\caption{Measurement duration in nights and hours by months of 2005 -- 2007 \label{tab:3}}
\bigskip
\centering
\begin{tabular}{l|rr|rr|rr}
\hline\hline
Season\ups &\multicolumn{2}{c}{2005} & \multicolumn{2}{|c|}{2006} & \multicolumn{2}{c}{2007}\\
Months & nights & hours & night & hours & night & hours \\[5pt]
\hline
January\ups &--- & --- & --- & --- &  1  &  0.1 \\
February  & ---  & --- & 8   &20.5 &  4  & 15.5 \\
March     & ---  & --- & 12  &27.2 &  6  &  8.8 \\
April   & ---  & --- & 17  &71.6 &  6  & 15.7 \\
May      & ---  & --- & 23  &84.1 & 10  & 23.5 \\
June     & ---  & --- & --- & --- &  2  &  4.8 \\
July     & ---  & --- & --- & --- & 26  & 72.5 \\
August   & 14   & 58.8& --- & --- & 25  & 87.8 \\
September & 30   &210.0& --- & --- & 21  &105.1 \\
October  & 23   & 73.8& 14  &15.9 & 20  & 61.4 \\
November   &  5   & 34.2&  5  & 9.6 &  7  & 18.5 \\
December  & ---  & --- & --- & --- & --- & ---  \\[5pt]
\hline\hline
\end{tabular}
\end{table}

Inspite of significant gaps in measurement periods, the observation time distribution follows more or less a typical seasonal clear nights allocation at Mt.~Maidanak except for December and January when, by various reasons, no scintillation was measured. The pronounced grouping of measurements in the five periods is well seen: fall 2005, spring and autumn 2006 and the ones of 2007.

\begin{figure}[!p]
\center
\psfig{figure=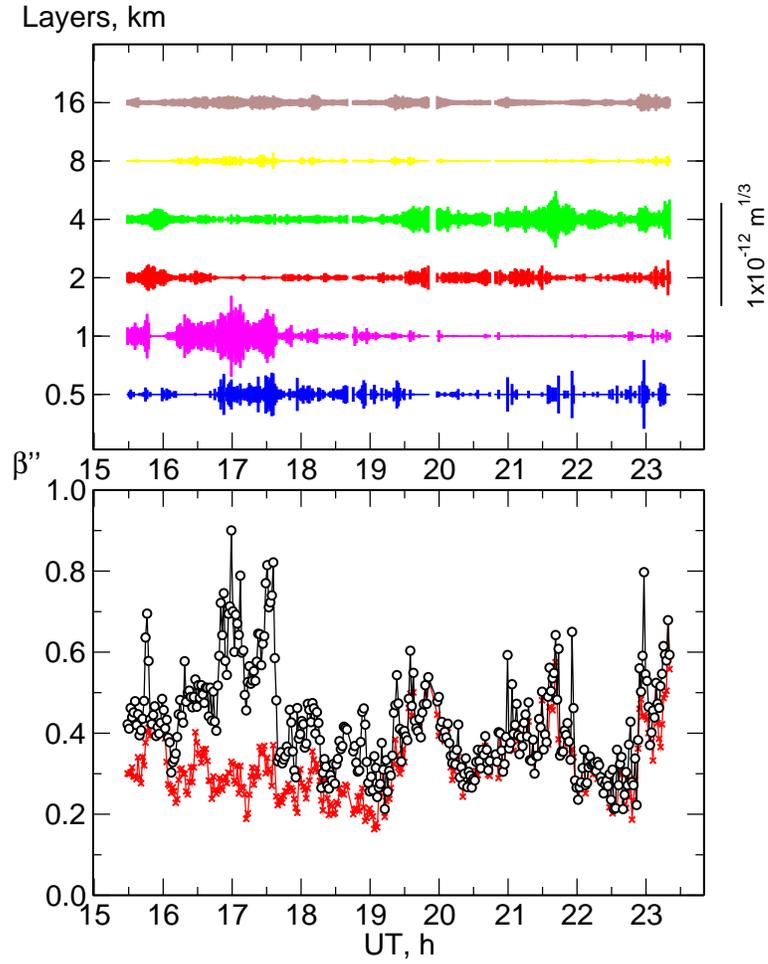,height=14cm}
\caption{ The September 9, 2005 results. Top: six fixed altitude layer representation of turbulence (see section~\ref{sec::mainres}). The stripe width is proportional to the layer power which scale is shown by the vertical bar to the right of the graph. Bottom: free atmosphere seeing $\beta_{free}$ behavior (circles) during the night; crosses depict seeing $\beta_2$ formed above 2~km. \label{fig:050906}}
\end{figure}

An example of MASS measurement results is shown in Fig.~\ref{fig:050906} for a typical night of September 9, 2005. The sporadic, usually quite short-term, increase of turbulence strength is seen in different layers (altitudes). The turbulence evolved from the evening time gradually declining in altitude. The end of night is dominated by the turbulence at 4~km.

During the campaign the Maidanak observatory differential image motion monitor (DIMM) was mainly used in other sites of Uzbekistan. Simultaneous measurements with both devices were made in 2005 from 17 of September to 5 of November (18 nights), in 2006 -- from 11 of February to 31 of May (24 nights) and continuously during 24 nights from 25 of August to 16 of September of 2007. The results of MASS and DIMM data inter-comparison will be presented in a separate paper.

\section{BASIC MEASUREMENT RESULTS}
\label{sec::mainres}
The MASS device is working under the {\it Turbina} program control \citep{MASS} which operates under Linux OS. Although the functionality of this program includes not only the measurement control but also the real-time data processing spanning from scintillation indices calculation to the restoration of a current altitude turbulence profile, the repetitive off-line data reprocessing allows one to obtain more reliable and homogeneous results. This improvement is related to temporal drifts of the device parameters over 2 -- 3 year period and, more importantly, to the continuous software update throughout the campaign time.

\begin{figure}[h]
\center
\psfig{figure=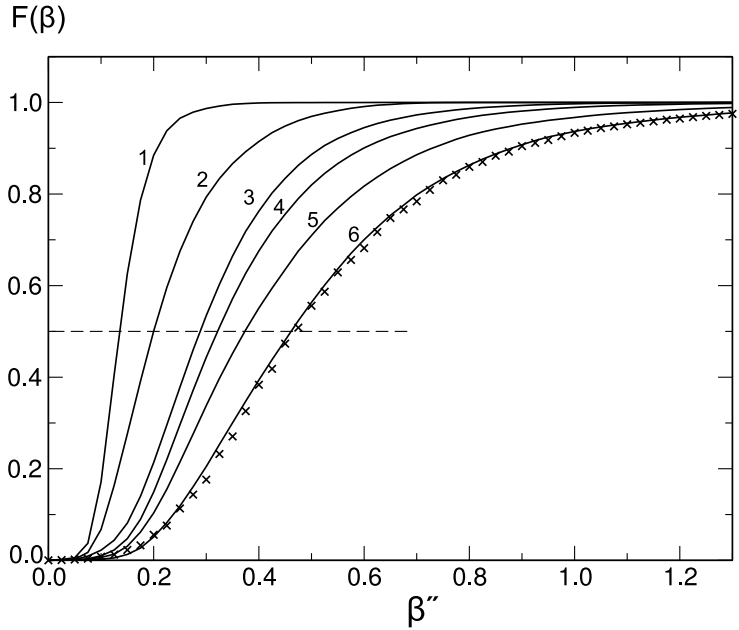,height=12cm,angle=-90}
\caption{ Cumulative distribution of the atmosphere layers input into the image quality: 1 --- 16~km layer, 2 --- 8~km layer and above, 3 --- 4~km and above, 4 --- 2~km and above, 5 --- 1~km and above, 6 --- all atmosphere from 0.5~km and above ($\beta_{free}$). Crosses --- $\beta_{free}$ distribution obtained directly from scintillation indices without profile restoration. \label{fig:cumm1}}
\end{figure}

Such a reprocessing\footnote{See Kornilov V., Shatsky N., {\it MASS Data Reprocessing}, \\ {\tt http://dragon.sai.msu.ru/mass/download/doc/remass.pdf, 2005.}} was performed with help of a special version of an {\it Atmos} program which is part of {\it Turbina} software, and a set of {\it shell}-scripts which filter out the input data glitches and invalid output. Although observations were performed with data accumulation time for a single profile restoration equal to 20~s due to imperfect tracking of AFR-2, the data reprocessing has regrouped the input into commonly accepted 1~min accumulation time.

As an outcome, more than 50 thousand integral atmosphere parameters sets and altitude profiles were computed. Each profile is quantified by turbulence intensities of six layers which are centered at 0.5, 1, 2, 4, 8 and 16~km. It should be stressed that provided values are not $C_n^2(h_i)$ but the integrated ones: $J_i =C_n^2(h_i)\Delta h_i$. Hence the input of each layer into the image quality $\beta$ is defined by the following relation:
$\beta_i = 2.0\times 10^{7}\cdot J_i^{3/5}$ for wavelength $\lambda = 500$~nm.
The seeing of $1\farcs0$ corresponds to the integral turbulence strength $J = 6.8\times 10^{-13}\mbox{m}^{1/3}$, and $\beta = 0\farcs5$ is produced by $J = 2.14\times 10^{-13}\mbox{m}^{1/3}$.

In Fig.~\ref{fig:cumm1} we show the cumulative distributions of the inputs of different atmosphere parts into the image quality. In other words, it shows the seeing for an observer at a given altitude above the site. Additionally the free atmosphere seeing $\beta_{free}$ distribution is shown, which was calculated as integral atmosphere parameter without profile restoration. One can see that these distributions are statistically indistinguishable.

The median layers contributions are given in Table~\ref{tab:4}. As a good comparison, the last column shows the median turbulence distribution for a Cerro Tololo observatory quoted from a paper by Tokovinin et al (\citeyear{CTIO03}).

\begin{table}
\caption{ Median contribution of atmosphere layers into the image quality (arcseconds). \label{tab:4}}
\bigskip
\centering
\begin{tabular}{r|r|rrrrr|r}
\hline\hline
Layer & Median $\beta$ & 2005A & 2006S & 2006A & 2007S & 2007A & CTIO'03\ups \\[5pt]
\hline
0.5~km and above & \ups 0.46 & 0.43 & 0.57 & 0.41 & 0.49 & 0.43 & 0.55\\
1~km and above   & 0.37 & 0.35 & 0.46 & 0.32 & 0.41 & 0.34 & --- \\
2~km and above   & 0.32 & 0.30 & 0.41 & 0.27 & 0.35 & 0.28 & 0.43\\
4~km and above   & 0.29 & 0.27 & 0.38 & 0.25 & 0.32 & 0.25 & 0.37\\
8~km and above   & 0.20 & 0.19 & 0.28 & 0.18 & 0.22 & 0.16 & 0.29\\
16~km and above  & 0.14 & 0.14 & 0.15 & 0.12 & 0.14 & 0.12 & 0.16 \\[5pt]
\hline\hline
\end{tabular}
\end{table}

An overall $\beta_{free}$ measurement median is $0\farcs46$. A somewhat smaller value of $0\farcs40$ was obtained for the Cerro Pachon observatory \citep{TokTrav06} in 2003 and 2005, while those value for Las Campanas site campaign is $0\farcs48$ \citep{LasCamp08}.

Interestingly, the formal median value precision for such a voluminous data set is only 0.01. Such an estimate of uncertainty is not adequate due to non-stationary nature of the considered phenomenon. The characteristic variation of a median seeing is illustrated by five seasonal medians.

\section{ISOPLANATIC ANGLE AND ATMOSPHERIC TIME CONSTANT} 

\begin{figure}[h]
\center
\psfig{figure=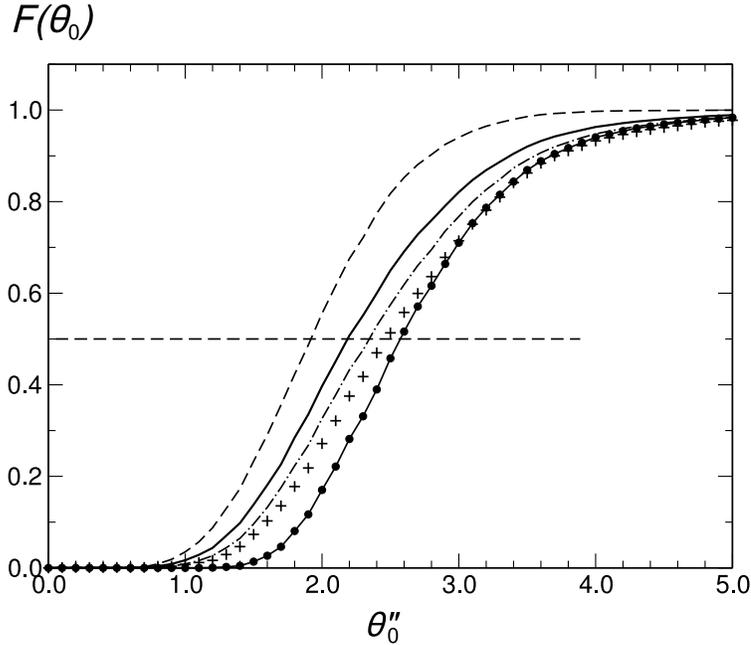,height=12cm,angle=-90}
\caption{ Cumulative isoplanatic angle  $\theta_0$ distribution for various seasons and conditions: solid line --- total scope distribution, dashed -- in springs, dot-dashed --- falls, crosses --- fall 2007, circles --- isoplanatic angle when $\beta_{free}$ is better than median.  \label{fig:cumm2}}
\end{figure}

Isoplanatic angle $\theta_0$ is involved to characterize a correlation of wavefront distortions in different directions \citep[see][]{Rodierbook}. It is computed from scintillation data as an integral parameter. Cumulative $\theta_0$ distribution is presented in Fig.~\ref{fig:cumm2} where the overall data distribution is accompanied by seasonal ones. The median overall measurement is $2\farcs19$ while it varies from $2\farcs35$ in fall to $1\farcs92$ in spring. Maximal median value was observed in autumn 2007  --- $2\farcs47$. As a reference it is worth to note the GSM estimate of 1998 -- 1999 \citep{eso} equal to $2\farcs48$ while Kornilov and Tokovinin (\citeyear{Free}) measured isoplanatic angle within $2\farcs12$ to $2\farcs40$ range.

In practice, the most interesting is a $\theta_0$ value in weak turbulence conditions. In Fig.~\ref{fig:cumm2} the $\theta_0$ distribution is shown for that half of measurements when seeing is less than median: $\beta_{free} < 0\farcs46$. The median of such a scope of $\theta_0$ is $2\farcs57$.

It is natural that $\theta_0$ is vastly determined by the upper atmosphere turbulence. In this respect its value should be typical for majority of sites located at 2.5~km elevation above the sea level \citep{CTIO03}. 

Equally important for adaptive optics performance, apart from isoplanatic angle, is an atmospheric time constant (widely also called a coherence time) $\tau_0$ which is another by-product of scintillation measurement \citep{time}. The dedicated study shows\footnote{See Tokovinin A., {\it Calibration of MASS time-constant measurements}, \\ {\tt http://www.ctio.noao.edu/$\sim$atokovin/profiler/timeconst.pdf}, 2006.} that the time constant from MASS estimation is systematically (factor of $\sim1.3$) lower than observed by other methods. Meanwhile we present uncorrected values hereafter.

\begin{figure}[h]
\center
\psfig{figure=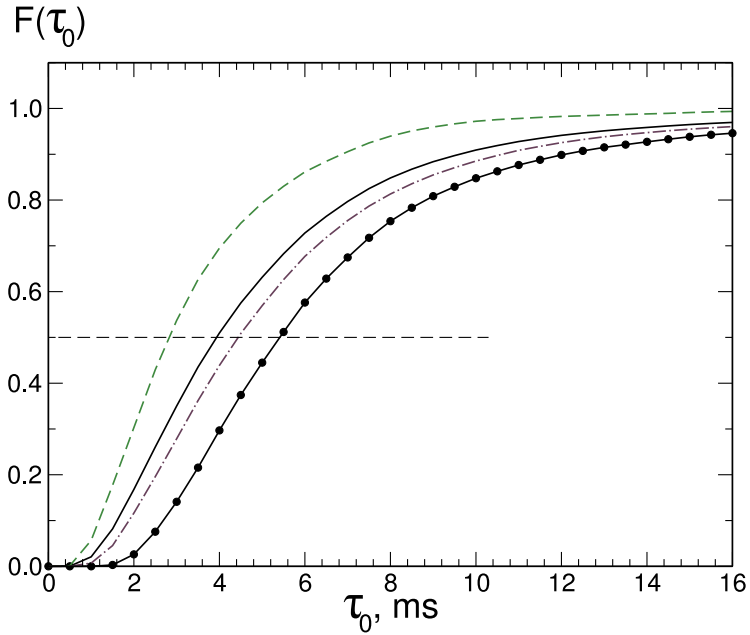,height=12cm,angle=-90}
\caption{ Cumulative distributions of atmospheric coherence time $\tau_{0}$: spring seasons --- dotted line, autumns --- dash-dotted, full data scope distribution --- solid. Black circles depict distribution when $\beta_{free}$ is better than median. \label{fig:cumm3}}
\end{figure}

Median of $\tau_0$ distribution shown in Fig.\ref{fig:cumm3} equals to 3.94~ms for the total campaign and 2.82~ms and 4.44~ms for springs and autumns, respectively. When restricted to $\beta_{free} < 0\farcs46$ conditions one obtains a 5.41~ms median. If we apply an empiric correction mentioned above, we come to the time constant about 7~ms in good conditions. The latter coincides with the median time constant at Antarctic plateau Dome~C obtained with MASS device in 2004 \citep{ken2006}.

One can recall the $\tau_0=12$~ms estimate of Ehgamberdiev et al (\citeyear{eso}) derived during the international astroclimate campaign of 1998 at Maidanak. Meanwhile, this value was based on a relatively short-term scope of measurements.

\section{PARTICULAR TURBULENCE PROPERTIES OF FREE ATMOSPHERE ABOVE MT. MAIDANAK}

\begin{figure}[h]
\center
\psfig{figure=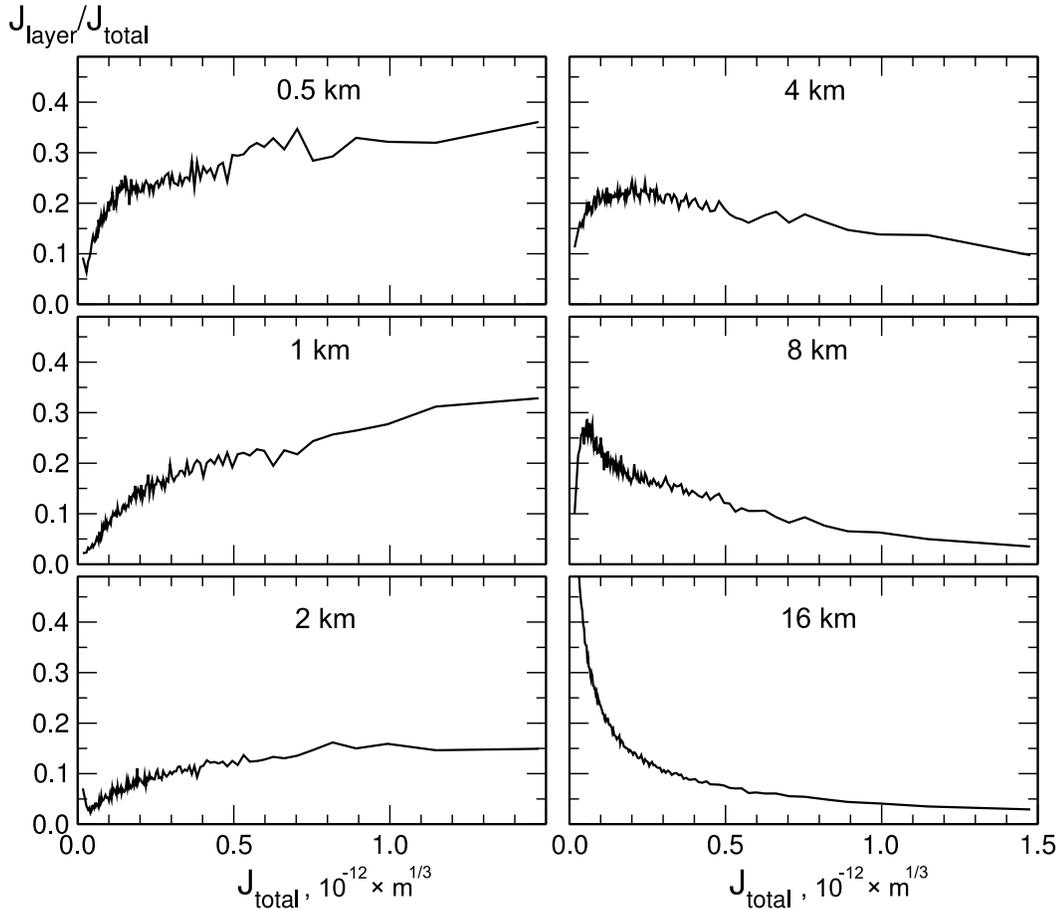,height=16cm,angle=-90}
\caption{ Relative input of atmosphere layers in a total turbulence as a function of latter \label{fig:frac}}
\end{figure}

In order to study the role of different atmosphere layers, we build the dependence of the relative layers intensity $J_{layer}/J_{total}$ on the total turbulence strength $J_{total}$ according to MASS measurements (i.e. excluding the ground layer). These ratios  are shown in Fig.~\ref{fig:frac} being averaged over 400 profiles. The layer input behavior differs strongly for the boundary layer (0.5 -- 1~km) and the upper atmosphere (8 -- 16~km). The former shows its input nearly proportional to the total power while the role of the upper atmosphere diminishes with the total growth of turbulence. Most representative here is the 16~km layer for which the relation is practically inversely proportional. This means that its intensity is almost invariable and absolutely insensitive to processes in the boundary layer. Similar phenomenon is noticed by Tokovinin and Travouillon (\citeyear{TokTrav06}) and in a paper by Thomas-Osip et al (\citeyear{LasCamp08}).

On the opposite, the boundary turbulence is responsible for image degradation at Maidanak. Keeping in mind the values $J = 0.2\times 10^{-12}\mbox{m}^{1/3}$ equivalent to $0\farcs5$ seeing, we are mostly interested in the graph beginnings up to values $0.5\times 10^{-12}\mbox{m}^{1/3}$. At the 2~km height the turbulence intensity is seemingly lower than at other altitudes.

\begin{figure}[h]
\center
\psfig{figure=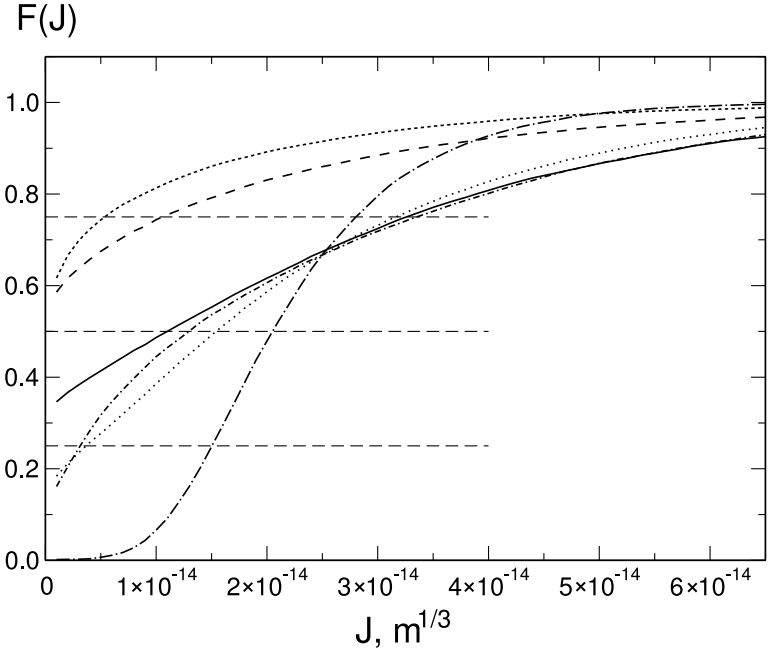,height=12cm,angle=-90}
\caption{ Cumulative distribution of turbulence intensity in 0.5~km layer (solid line), at 1~km  (short dashes), 2~km (long dashes), 4~km (dots),  8~km (short dash-dots) and 16~km (long dash-dots) for cases of good images ($\beta_{free}<0\farcs46$).  \label{fig:power}}
\end{figure}

As a support for this fact of a key role of the boundary layer, we present the Fig.~\ref{fig:power} where the cumulative distributions of layers intensities are shown for a good seeing data subset (image quality better than median). The prominent feature is a nearly absent lower (1 and 2~km) turbulence for 50\% of cases. The median of intensity in  4- and 8~km layers is $1.3\div 1.5\times 10^{-14}\mbox{m}^{1/3}$ which is slightly more than the median value for the lowest 0.5~km layer ($1.1\times 10^{-14}\mbox{m}^{1/3}$). It is natural to think that at 0.5~km we see the development of the turbulence generated by a ground relief.

The special case here is again the upper atmosphere at 16~km. The steep rise of the cumulative distribution tells about nearly constant turbulence which is quite significant. In half a time its intensity is in a narrow span of $1.5\div 2.8\times 10^{-14}\mbox{m}^{1/3}$. Note that the high altitude turbulence is restored with a minimal error from the scintillation data.

\section{CONCLUSIONS}

An estimation of a total turbulence power from DIMM measurements \citep{I99} allows us to evaluate the dominating input of the ground turbulence (up to 500~m above surface) to be 65\% on average and the first 200 -- 300~m above surface give already about 50\% contribution. The same contribution is found at Cerro Pachon by Tokovinin and Travouillon (\citeyear{TokTrav06}).

In cases of exceptionally good images (25\% quantile) the full image quality from DIMM measurements in 1996 to 1999 \citep{eso} constitutes $0\farcs55$. Comparing with data in Fig.~\ref{fig:cumm1} we come to the ground layer contribution of 60\% in these conditions while the whole boundary layer inputs more than 70\% leaving less than one third of a power to the rest of atmosphere.

In such a situation the development of an adaptive optics system working in the visible should be directed to correction of the low turbulence. It is hardly possible to achieve the Strehl ratio improvement factor more than 5 here, meanwhile the large corrected AO field of view \citep[of the order of several arc-minutes, see][]{AOFOV} will be a reasonable compensation. It is worth to note here that the maximum gain for the median Maidanak conditions (Fried radius $r_0=0.15\mbox{m}$ at wavelength 500~nm) at the 1.5~m telescope AZT-22 computed according to Roddier formula (\citeyear{R1998}) for optimal atmospheric distortions correction is equal to 32.

A profiting advantage of Maidanak observatory is a large time constant value $\tau_0 \approx 7\mbox{ms}$. This lowers the critical demand of a system response time by a factor of nearly 2.5 and thus makes about three times more stars available for wavefront referencing compared to Cerro Tololo and Cerro Pachon observatories \citep{ken2006}. Such a good $\tau_0$ also favours the development of optical interferometry at this site.

This research was supported by Russian Basic Research Foundation (grant 06-02-16902a). The staff of Maidanak observatory helped a lot in organization and conduction of measurements for which we express gratitude to all of them.

\pagebreak   

\end{document}